\documentclass[twocolumn,preprintnumbers,amsmath,amssymb]{revtex4}

\usepackage{graphicx}
\usepackage{braket}

\newcommand{\beginmethods}{%
        \setcounter{table}{0}
        \renewcommand{\thetable}{M\arabic{table}}%
        \setcounter{figure}{0}
        \renewcommand{\thefigure}{M\arabic{figure}}%
     }

\newcommand{\beginsupplement}{%
        \setcounter{table}{0}
        \renewcommand{\thetable}{S\arabic{table}}%
        \setcounter{figure}{0}
        \renewcommand{\thefigure}{S\arabic{figure}}%
     }

\begin{document}

\title{Testing universality of Efimov physics across broad and narrow Feshbach resonances}
\author{Jacob Johansen}
\author{B. J. DeSalvo}
\author{Krutik Patel}
\author{Cheng Chin}
\affiliation{The James Franck Institute, Enrico Fermi Institute, and Department of Physics, \\ The University of Chicago, Chicago, IL 60637, USA}

\begin{abstract}
Efimov physics is a universal phenomenon arising in quantum three-body systems. For systems with resonant two-body interactions, Efimov predicted an infinite series of three-body bound states with geometric scaling symmetry \cite{Efimov1}. These Efimov states were first observed in cold Cs atoms \cite{Grimm1} and have since been reported in a variety of atomic systems \cite{Hulet1,Zaccanti1,Gross1,Jochim2,OHara1,Jin1,LENS1,Grimm2,LiRb,Chin2,Weidemuller3,HeEfimov}. While theories predict non-universal behavior for narrow Feshbach resonances \cite{Petrov,Gogolin,Schmidt2012,Aarhus1}, experiments on Efimov resonances are so far consistent with predictions based on universal theories. Here we directly compare the Efimov spectra in a $^6$Li-$^{133}$Cs mixture near two Feshbach resonances which are very different in their resonance strengths but otherwise almost identical. Our result shows a clear dependence of the Efimov resonance positions on Feshbach resonance strength and a clear departure from the universal prediction for the narrow Feshbach resonance.
\end{abstract}

%\pacs{34.50.Cx}

\maketitle

%The three-body parameter is crucial to understanding Efimov physics, as it determines the absolute positions of all Efimov resonances. Previous experiments in homonuclear systems yield an interesting observation: the first Efimov resonance position $a_-$ appears to follow a universal formula $a_-\approx -9.5~r_{vdW}$ \cite{Grimm3}, where $r_{vdW}$ is the van der Waals length of the molecular potential. Calculations from universal theory based on a single channel model confirm this ``van der Waals universality'' for Feshbach resonances with large resonance strength $s_{\mathrm{res}}>1$ \cite{Chin3,Greene1,Schmidt2012}. Similar theories also give universal but modified predictions for heteronuclear systems \cite{Esry2,Greene3}. For resonances with $s_{\mathrm{res}}\ll1$, however, significant deviations from the universal prediction are expected \cite{Petrov,Gogolin,Schmidt2012,Aarhus1}. On the other hand, previous experiments reaching down to $s_\mathrm{res}=0.11$ have shown little or no dependence on $s_{\mathrm{res}}$ \cite{LENS1}. Experimental results, including this work, are summarized in Fig. \ref{fig0}.

\begin{figure}[h]
\includegraphics[clip,trim=0 0 0 .15in,width=3.5 in]{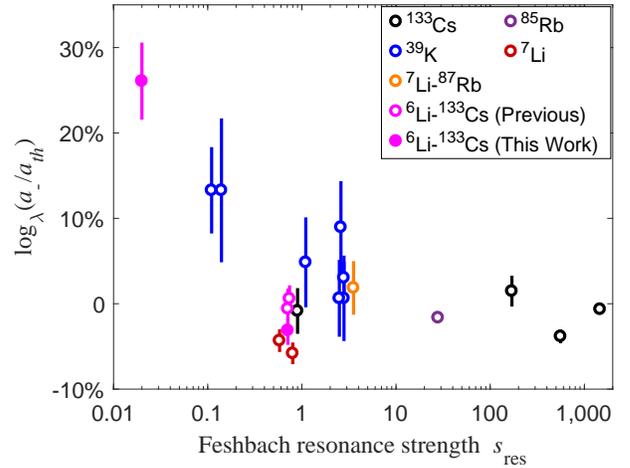}
\caption{Deviation of measured first Efimov resonance positions $a_-$ from predictions based on universal theory $a_{th}$. For identical bosons, $a_{th}=-9.73~r_\mathrm{vdW}$ \cite{Greene1}. For the $^7$Li-$^{87}$Rb Efimov resonance, the universal prediction yields $a_{th}=-1,800~a_0$ \cite{LiRb}. In $^6$Li-$^{133}$Cs, $a_{th}=-320~a_0$ for the 843 G resonance, $a_{th}=-2,150~a_0$ for the 889 G resonance \cite{Weidemuller2}, and $a_{th}=-2,200~a_0$ for the 893 G resonance \cite{Yujun1}. To compare systems with different Efimov periods $\lambda$, fractional deviations are shown. Previous measurements are plotted as open circles: $^{133}$Cs \cite{Grimm3} (black), $^7$Li\cite{Hulet1,Gross1} (red), $^{39}$K \cite{LENS1} (blue), $^{85}$Rb \cite{Jin1} (purple), $^7$Li-$^{87}$Rb \cite{LiRb} (orange), and $^6$Li-$^{133}$Cs \cite{Chin2,Weidemuller3,Weidemuller2} (magenta). Data from this work are shown as closed circles. Error bars correspond to the one standard deviation total uncertainty.} \label{fig0}
\end{figure}

The existence of Efimov states is evidenced by the observation of Efimov resonances in a variety of ultracold atomic systems \cite{Hulet1,Zaccanti1,Gross1,Jochim2,OHara1,Jin1,LENS1,Grimm2,LiRb,Chin2,Weidemuller3}. Experiments in homonuclear systems lead to an interesting observation: the first Efimov resonance position $a_-$ appears to follow a universal formula $a_-\approx -9~r_\mathrm{vdW}$ \cite{Grimm3}, where $r_\mathrm{vdW}$ is the van der Waals length of the molecular potential. Calculations from universal theory based on a single scattering channel model confirm this ``van der Waals universality'' for Feshbach resonances with large resonance strength $s_{\mathrm{res}}>1$ \cite{Chin3,Greene1,Schmidt2012,Ueda1,Ueda2}. Similar theories also give universal but modified predictions for heteronuclear systems \cite{Esry2,Greene3}. For resonances with $s_{\mathrm{res}}\ll1$, significant deviations from the universal theory are expected \cite{Petrov,Gogolin,Schmidt2012,Aarhus1}. Previous experiments reaching down to $s_\mathrm{res}=0.11$, however, have shown little or no dependence on $s_{\mathrm{res}}$ \cite{LENS1}. Experimental results, including this work, are summarized in Fig. \ref{fig0} and the Supplement.

% For heteronuclear systems, the universal prediction depends on the mass ratio and intraspecies scattering length.

\begin{figure}[t]
\includegraphics[width=3.5 in]{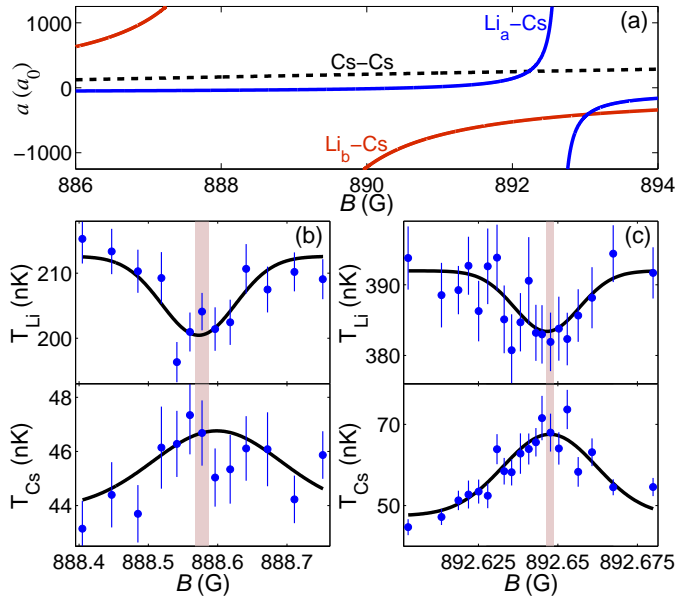}
\caption{Feshbach resonances employed in this work. (a) Calculated scattering lengths as a function of magnetic field in the region of interest, showing a narrow resonance in the Li$_\textrm{a}$-Cs spin channel (blue line) and a broad resonance in the Li$_\textrm{b}$-Cs spin channel (red line) \cite{Chin4}. The Cs-Cs scattering length is shown as a dashed line. In (b) and (c), we show cross-thermalization between Li and Cs across the Feshbach resonances after interaction times of 200 and 250 ms, respectively. Gaussian fits yield positions of 888.577(10)(10) and 892.648(1)(10) G, respectively. The shaded region on each plot indicates the one standard deviation statistical uncertainty in Feshbach resonance position, determined from the variance weighted average of the temperature fits (black curves).} \label{fig1}
\end{figure}

In this Letter, we compare Efimov spectra in a $^6$Li-$^{133}$Cs mixture near one broad ($s_\mathrm{res}=0.71$) and one very narrow ($s_\mathrm{res}=0.02$) interspecies Feshbach resonance at $B_0=889$ and 893 G, respectively. These two Feshbach resonances, shown in Fig. \ref{fig1}(a), differ significantly in resonance strength, but only slightly in Cs-Cs scattering length ($a_\mathrm{CsCs}=200$ and $260~a_0$, where $a_0$ is the Bohr radius). Therefore, this pair of resonances is an excellent case to assess the influence of the Feshbach resonance strength on Efimov physics while keeping other parameters nearly identical. We observe a significant difference in Efimov resonance position between the two Feshbach resonances. Our measurements also show that the resonance associated with the first Efimov state is suppressed near both Feshbach resonances. %and see suppression of three-body loss near the narrow resonance at positive scattering length.

To resolve Efimov features near the narrow Feshbach resonance at 893 G, we require milligauss level control of our magnetic field. For this, we improve our magnetic field control from our previous work \cite{Chin2}. In addition, we develop a scheme to precisely monitor the magnetic field, achieving a precision below 3 mG in each measurement (see Methods). We also incorporate a new dual color optical trap to eliminate relative gravitational sag of the atoms, which prevented mixing of the two species at temperatures below 200 nK in our previous work \cite{Chin2}, see Methods.% The new optical trap combines 1064 and 785 nm lasers, allowing complete overlap of Li and Cs clouds at low temperatures (see Methods).

With these improvements to the apparatus, we prepare spin polarized Li and Cs samples at temperatures as low as 50 nK with tunable overlap. We vary magnetic field to tune the interspecies scattering length. In each scan, we randomly order the magnetic fields to eliminate systematic drifts; we also choose an interaction time and overlap to optimize the sensitivity to the collision rate. Details regarding the experimental preparation are provided in the Methods.

%Atoms are initially prepared at a magnetic field far from resonance where Li-Cs interactions are negligible, and then B is quickly ramped to a field of interest.  To avoid systematic drifts in our data, we randomly order the magnetic fields inverstigated.  Atoms are the held for an interaction time chosen to enhance sensitivity to the collision rate.  After the interaction time, B is ramped back to a value with negligible Li-Cs interactions and the atoms are imaged.

To determine the interspecies scattering length $a$ at which features occur, we require precise knowledge of Feshbach resonance positions. Previous work based on three-body loss and radio-frequency spectroscopy \cite{Chin4,Weidemuller1} offers a model to extract scattering length from the magnetic field; however, higher magnetic field precision is needed for analysis near the narrow resonance. We develop a procedure to precisely determine the resonance positions based on cross-species thermalization. We first create a significant temperature difference between Li and Cs samples, with $T_\mathrm{Cs}\approx$45 nK and $T_\mathrm{Li}\approx$300 nK. After the interaction time, we measure the final temperatures. Near an interspecies Feshbach resonance, enhanced elastic collisions between the two species lead to enhanced interspecies thermalization. This approach has a significant advantage over measurements relying on three-body recombination or radio-frequency spectroscopy because the elastic collision rate is symmetric about the resonance over the small range we probe and does not depend on a complex model to extract the resonance position \cite{Chin1}. %This technique has a significant advantage over former measurements which require theoretical modeling and interpretation of asymmetric lineshapes, as the elastic collision rate is symmetric about the resonance \cite{Chin1}. %

%In addition, we only partially overlap the lithium and cesium clouds. Partial overlap suppresses the three body loss rate relative to the thermalization rate and ensures that magnetic field settling time is negligible relative to the interaction time. For this measurement, the high lithium temperature also ensures that our lithium does not form a degenerate Fermi gas and introduce potential many-body effects. Methods section

The interspecies thermalization measurements are shown in Fig. \ref{fig1}. We determine the resonance positions with a Gaussian fit to the measured temperatures. The extracted resonance positions from Li and Cs data agree within the uncertainty. The variance weighted resonance positions are $B_\mathrm{broad}=888.577$(10)(10) G and $B_\mathrm{narrow}=892.648$(1)(10) G, where the values in parentheses indicate the statistical and systematic uncertainties, respectively. Notably, the systematic uncertainty is dominated by factors which are common to all measurements. As a result, magnetic field offsets relative to the Feshbach resonance positions, $\Delta B_\mathrm{b}$ and $\Delta B_\mathrm{n}$, have a reduced systematic uncertainty (see Methods). Our results are consistent with Refs. \cite{Chin4,Weidemuller1} while offering much higher precision.

\begin{figure*}[t]
\includegraphics[clip,trim=0 1.05in 0 1.2in,width=\textwidth]{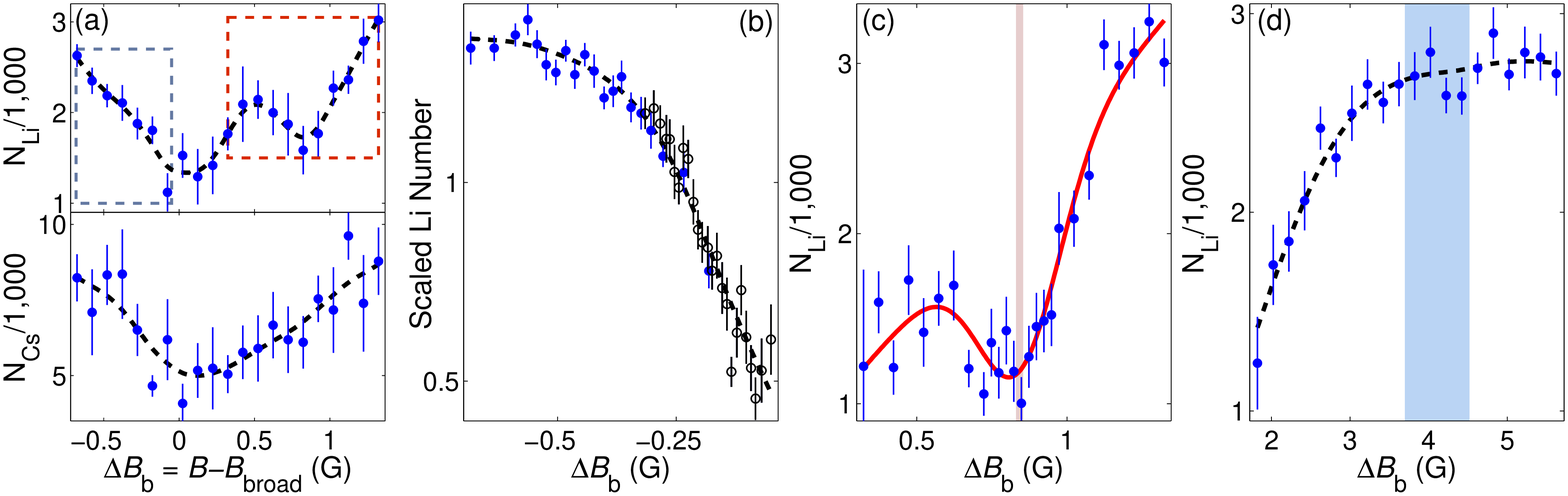}

\includegraphics[clip,trim=0 1.05in 0 1.2in,width=\textwidth]{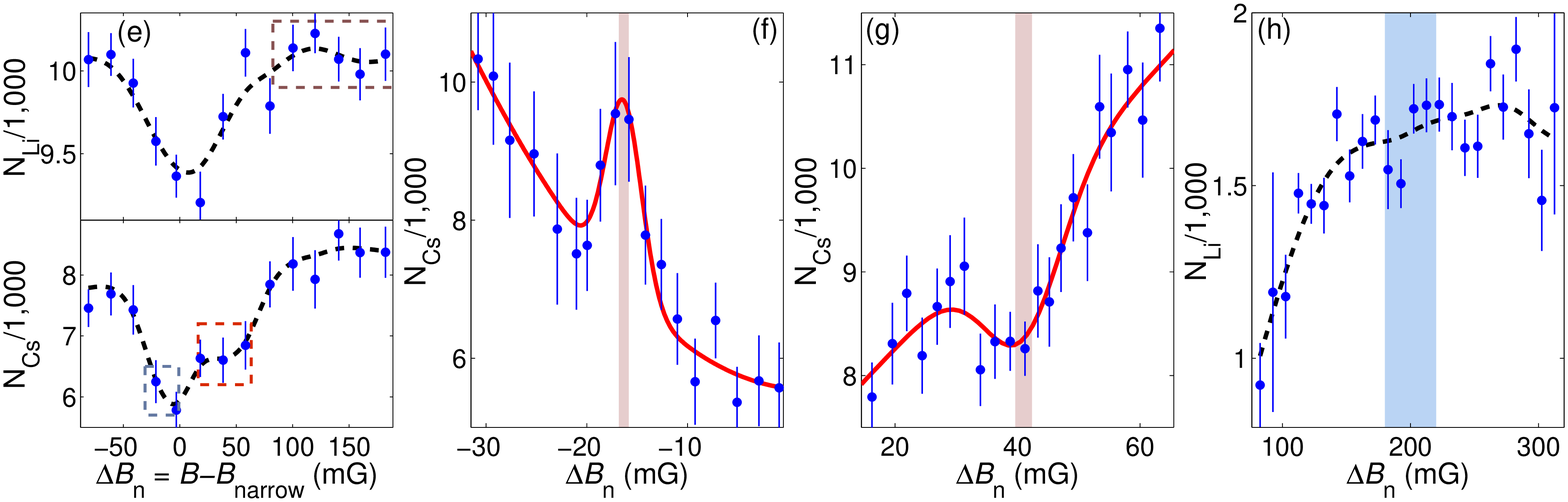}
\caption{Measurement of Efimov features near the broad and narrow Feshbach resonances. All magnetic fields are plotted relative to the Feshbach resonance positions determined in Fig. \ref{fig1}. Panels (a)--(d) show the features near the broad resonance, while (e)--(h) show the features near the narrow resonance. Panels (a) and (e) show broad scans across the Feshbach resonances. In panels (a) and (e), gray, red, and brown boxes indicate the regions of the fine scans shown in (b) and (f), (c) and (g), and (h), respectively. Panels (b) and (f) are on the positive scattering length side of Feshbach resonances. Near the narrow resonance, we identify suppression of three-body loss at $\Delta B_\textrm{n}$=-16(1)(2) mG. Scaled atom number is shown in (b) to connect scans performed under different conditions. We observe an Efimov resonance on the negative side of each Feshbach resonance, shown in (c) at $\Delta B_\textrm{b}$=842(12)(10) mG and in (g) at $\Delta B_\textrm{n}$=41(1)(2) mG. Panels (d) and (h) show scans in the range expected for the lower order Efimov resonance; the blue shaded regions indicate the expected positions assuming a scaling constant within $\pm$10\% of $\lambda=4.88$ \cite{Greene3}. Black dashed curves are guides to the eye. Red solid curves are fits used to extract positions of features, while the corresponding red shaded regions indicate one standard deviation statistical error bars.}%The interaction time for each scan is: (a) 80 ms, (b) 400 (closed circles) and 200 ms (open circles), (c) 80 ms, (d) 1.2 s, (e) 80 ms, (f) 250 ms, (g) 80 ms, (h) 700 ms}
\label{fig2}
\end{figure*}

To see Efimov features near these Feshbach resonances, we modify the experimental sequence to optimize the three-body loss signal. We prepare Li and Cs at nearly equal temperatures of $\approx$100 nK. After the interaction time, we measure Li and Cs atom numbers by absorption imaging (see Methods).

The atomic loss spectra are shown in Fig. \ref{fig2}. Coarse scans across the Feshbach resonances provide context for the Efimov features, see Figs. \ref{fig2}(a) and (e). Finer scans at negative scattering length, shown in Figs. \ref{fig2}(c) and (g), display enhanced loss features, identified as Efimov resonances. These features are qualitatively similar, but with different magnetic field scales. Using a Gaussian fit with a linear background, we determine the Efimov resonance positions to be 0.842(12)(10) and 0.041(1)(2) G above the broad and narrow Feshbach resonance, respectively. Using the model in Ref. \cite{Weidemuller1} and our measurements of the Feshbach resonance positions, we calculate the corresponding scattering lengths as $a=-$2,050(60) and $-$3,330(240)$~a_0$ for the broad and narrow resonance, respectively. The quoted uncertainties in scattering length include the previously discussed statistical and systematic uncertainties in magnetic field, as well as an additional 2\% uncertainty in the model \cite{Weidemuller1}.

Comparing with our previous measurement near the broad Feshbach resonance at 843 G \cite{Chin2}, where the first Efimov resonance occurs at $-323(8)~a_0$, both Efimov resonance positions in the present work are more than a full Efimov period higher in magnitude. This suggests the existence of a lower order Efimov state. However, searching in the ranges expected from geometric scaling \cite{Chin2}, we do not observe any additional Efimov resonances, see Figs. \ref{fig2}(d) and (h). The missing feature has been studied for the resonance at 889 G \cite{Weidemuller2} and arises from a weakly bound Cs$_2$ state, which suppresses the resonance due to the ground Efimov state. We find that the first Efimov resonance is missing for both the broad and narrow Feshbach resonances measured here. Based on this interpretation, the Efimov resonances observed in this work stem from the second Efimov states.

\begin{figure}[h]
\includegraphics[clip,trim=0 .6in 0 .9in,width=3.5 in]{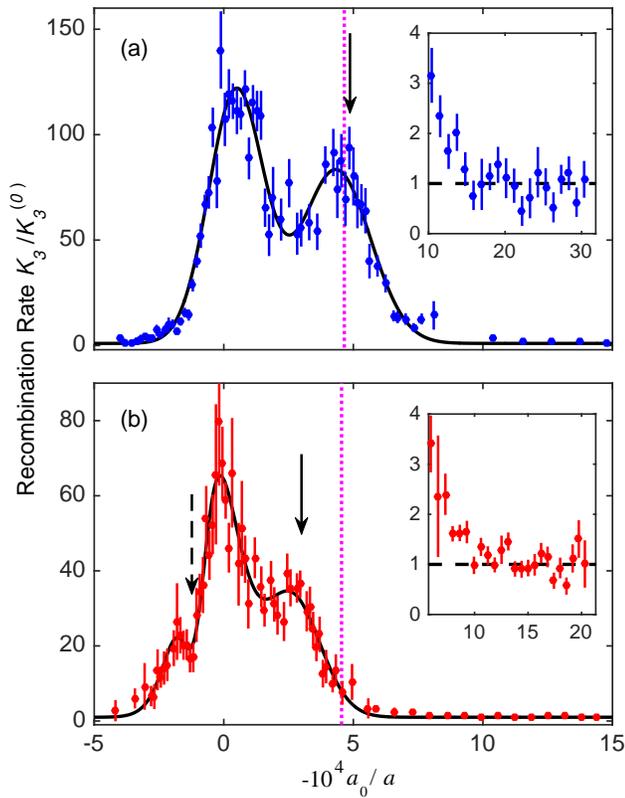}
\caption{Recombination rate $K_3$ as a function of inverse scattering length for the broad resonance (a) and narrow resonance (b) near 890 G. The recombination rate is normalized to the off-resonant value $K^{(0)}_3$, see dashed lines in the insets. Vertical solid arrows indicate Efimov resonance positions. The dashed arrow in (b) indicates a suppression feature in the three-body loss rate at positive scattering length near the narrow resonance. The magenta dotted lines indicate the positions predicted by the universal theory.} \label{fig3}
\end{figure}

Additionally, we observe suppression of three-body loss at positive interspecies scattering length, see Fig. \ref{fig2}(f) and Supplement. The suppression occurs at $\Delta B_\textrm{n}=-16$(1)(2) mG, corresponding to $a=8$,100(1,300)$~a_0$. Searching for a similar feature near the broad Feshbach resonance over the range 887.9 to 888.5 G ($a=$2,500 to 30,000$~a_0$), however, yields no discernible features, see Fig. \ref{fig2}(b).%Also note that, while the rest of the scans in Figure \ref{fig2} show the absolute atom number, we show a scaled atom number in 2(b). This is because these scans are performed under slightly different conditions in order to maximize the signal to noise ratio for each scan, and therefore cannot be directly compared to one another in terms of absolute number.

We present the results of all scans in terms of interspecies scattering length in Fig. \ref{fig3} to compare Efimov features near the broad and narrow Feshbach resonances. To combine all scans, we convert atom number to the three-body recombination coefficient $K_3$ based on a numerical model of three-body loss described in the Methods. Owing to uncertainties in trapping frequencies and overlap between the Li and Cs clouds, we scale $K_3$ to the measured value far from resonance $K^{(0)}_3$, shown in Fig. \ref{fig3} insets. From this figure, we can clearly see a difference in the Efimov resonance positions for the broad and narrow Feshbach resonances, indicated by the solid arrows. We can also see significant deviation from the universal prediction (magenta dotted lines) for the narrow Feshbach resonance.

For precise quantitative comparison of Efimov resonances, as well as comparison to the universal theory, we summarize Feshbach and Efimov resonance positions in $^6$Li-$^{133}$Cs in Table \ref{table1}. Due to the suppressed resonance from the first Efimov states, we compare resonance positions from the second Efimov states $a^{(2)}_-$. Between the broad and narrow Feshbach resonances near 890 G, we observe a large difference of 63(12)\% in $a^{(2)}_-$, while the universal theory predicts a much smaller effect of 2.3\% associated with the small difference in $a_\mathrm{CsCs}$. Furthermore, the universal prediction is in excellent agreement with measurements for both broad resonances at 843 and 889 G, in spite of the significant difference in $a_\mathrm{CsCs}$. This agreement indicates that the universal theory describes broad resonances. However, it does not capture the behavior of the narrow resonance, as our measurement near the narrow resonance at 893 G deviates from the universal theory by 51(11)\%.

\begin{table}[h]
\caption{Summary of Efimov resonances for the three Feshbach resonances in Li-Cs at 843 \cite{Chin2}, 889, and 893 G. Here, $B_0$ indicates the Feshbach resonance position, $s_\mathrm{res}$ the Feshbach resonance strength, $a_\mathrm{CsCs}$ the Cs-Cs scattering length, $a^{(2)}_-$ the resonance position associated with the second Efimov state, and $a^{(2)}_{th}$ the prediction from the universal theory using a single channel model \cite{Weidemuller2,Yujun1}.}
\begin{center}
\begin{tabular}{c c r r r r} 
\hline
$B_{0}$(G) & $\, s_\mathrm{res}\, $ & $a_{\mathrm{CsCs}}$($a_0$) & $a^{(2)}_{-}$ ($a_0$) & $ a^{(2)}_{th}$ ($a_0$)\\
\hline
888.577(10)(10) & 0.71 & $200$ & $-2,050$(60) & $-2,150$\\
892.648(1)(10) & 0.02 & $260$ & $-3,330$(240) & $-2,200$\\
842.75(1)(3) & 0.74 & $-1,400$ & $-1,635$(60) & $-1,680$\\
\hline
\end{tabular}
\end{center}
\label{table1}
\end{table}

Placing this work in the broader context of previous studies offers insight into the universality of Efimov  physics in cold atom systems. Based on the general form $K_3\propto \left\{\mathrm{sin}^2[\pi\mathrm{log}_\lambda (a/a_-)]+\mathrm{const.}\right\}^{-1}$ for $a<0$ \cite{Esry4}, we introduce a phase shift $\Delta\Phi=\pi\mathrm{log}_\lambda (a_-/a_{th})$ which characterizes deviation from the universal prediction and accounts for the scaling constant $\lambda$. As such, the metric $\Delta\Phi/\pi$ can be used to compare our result to other experiments with different scaling constants, as shown in Fig. \ref{fig0}. Our result at $s_\mathrm{res}=0.02$, the smallest resonance strength thus far investigated, offers a clear deviation of 26(5)\% of an Efimov period from the universal prediction, a 5 standard deviation effect. As published theories describing narrow resonances are constrained to homonuclear systems, deeper understanding of the Efimov resonance dependence on $s_\mathrm{res}$ will require theoretical work tailored to heteronuclear systems beyond single channel calculations. Nevertheless, our result already suggests a trend of increasing Efimov resonance position with decreasing $s_\mathrm{res}$.

We thank Yujun Wang, Chris Greene, and Rudolf Grimm for valuable discussion. We thank Lei Feng, Colin V. Parker, Karina Jim\'{e}nez-Garc\'{\i}a, and Shih-Kuang Tung for help in the early stages of the experiment. We acknowledge funding support from NSF Materials Research Science and Engineering Centers grant DMR-1420709 and NSF grant PHY-1511696. B. J. D. is supported by the Grainger Fellowship.

\section*{Methods}

\beginmethods

\noindent\textbf{Dual color optical dipole trap.} In our current work, we use an updated optical dipole trapping scheme. This includes the 1064 nm elliptical trap and $\sim1070$ nm translatable trap discussed in previous work \cite{Chin4,Chin2} with some minor changes. The elliptical trap is approximately 33 by 350 $\mu$m at the atomic position, with the tight axis along the vertical direction. As before, it may be translated vertically or modulated to increase its vertical size \cite{Chin2}. However, as in previous work, gravitational sag separates Cs from Li in this trap at very low temperatures.

To circumvent this, we implement a dual color trap consisting of 1064 and 785 nm light co-propagating with the elliptical beam, with maximum powers of 150 and 120 mW, respectively. In our present work, both beams are focused to a waist of 29 $\mu$m with the 1064 nm beam approximately 5 $\mu$m above the 785 nm beam. The 785 nm beam pushes Cs up while pulling down Li, allowing us to counteract the difference in gravitational sag. Using this trap, we may maintain perfect overlap at arbitrarily low temperatures by careful choice of the 785, 1064, and elliptical beam powers, as well as the elliptical beam position.

As all of these beams are co-propagating, they only very weakly trap atoms along the propagation (axial) direction. While an additional beam is used to provide axial confinement during initial loading, in the final trap this beam is turned off, and axial confinement is provided by magnetic field curvature with a trapping frequency of approximately 6.5 Hz for Cs and 36 Hz for Li near 900 G. The use of magnetic trapping ensures that the atoms are located at the center of curvature of our magnetic field, allowing us to better measure our magnetic field with only small adjustments to our experimental procedure, as detailed below.

\begin{figure*}[t]
\includegraphics[clip,trim=.5in .6in 0 .3in,scale=.52]{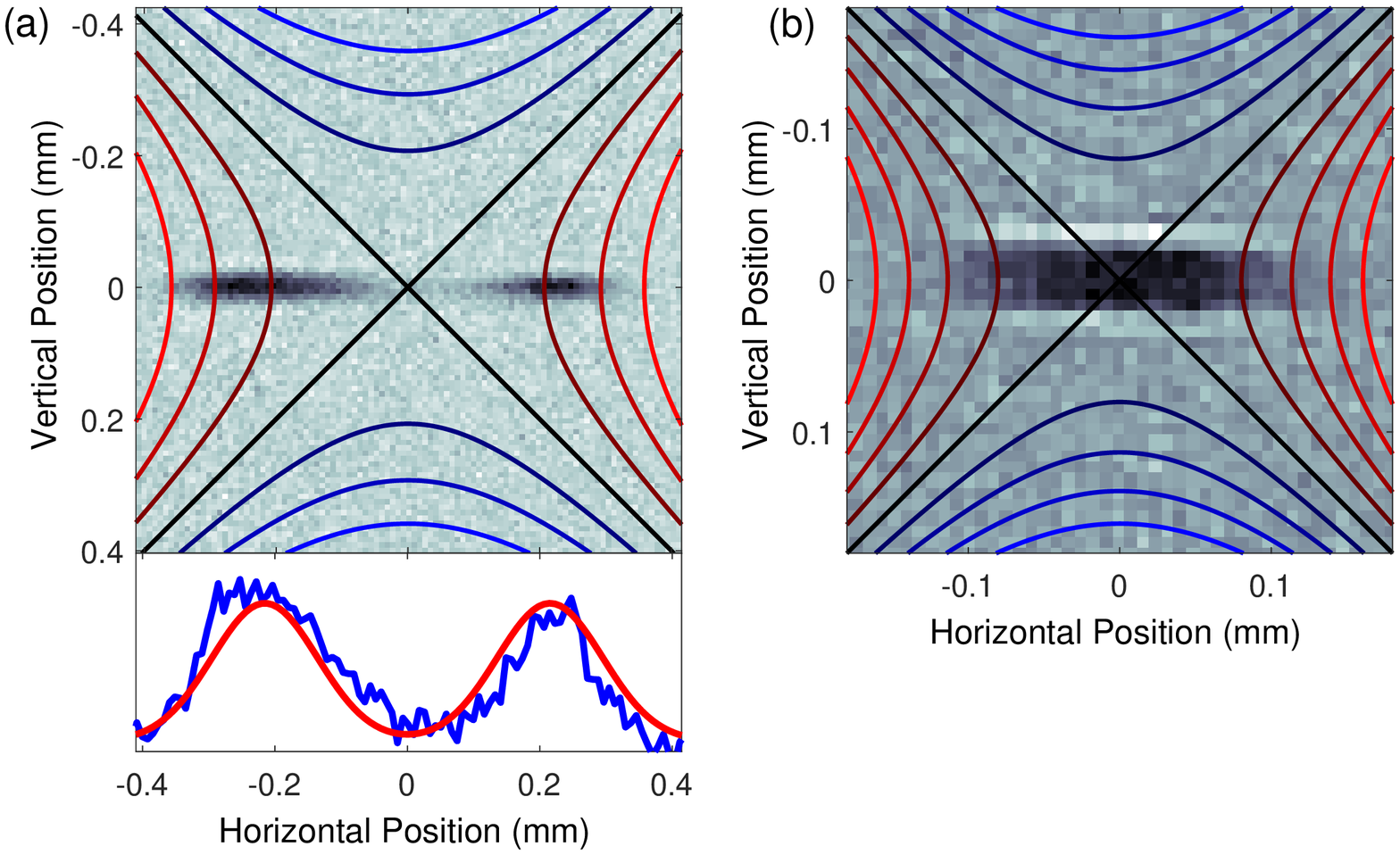}
\caption{Tomographic approach to magnetic field calibration. (a) Typical \textit{in situ} Cs image used for field calibration, with contours indicating constant magnetic field, spaced by 20 mG: Red indicates smaller magnetic field; blue larger field. In this image, we detune the microwave frequency such that it selectively excites atoms in two regions where the magnetic field is approximately 25 mG below the value at the center. The precise field is determined by fitting the distance between the two regions, as shown in the lower panel. (b) Colder sample ($\approx100$ nK), with magnetic field contours spaced by 3 mG. The width of this cloud corresponds to a spread in magnetic field of approximately 2 mG, as can be seen by the fact that nearly all atoms are within the area enclosed by the first contour.}
\label{figMeth1}
\end{figure*}

\noindent\textbf{Magnetic field calibration.} As in our previous work, we use the $\ket{F,m_F}=\ket{3,3}$ to $\ket{4,4}$ (where $F$ and $m_F$ denote the total angular momentum and magnetic quantum number, respectively) microwave transition in cesium to calibrate our magnetic field $B$. However, we now use a tomographic approach to measure the field in a single execution of the experimental sequence. We use our ability to move the elliptical beam to measure the field curvature to a precision of $\approx 95$\%. To measure the field, we run a modified version of the sequence, omitting most of the cesium evaporation and leaving cesium at a temperature of 1--2 $\mu$K. We then use a 1 ms microwave pulse set to a frequency slightly below that which would pump the center of the cloud, and immediately image the atoms in the $\ket{4,4}$ state. Because of the magnetic field curvature, two regions of the cloud are imaged, as shown in Fig. \ref{figMeth1}(a). The separation between these two imaged regions is determined by the magnetic field curvature, the microwave frequency, and the magnetic field at the center of the cloud. Because we use magnetic trapping to provide axial confinement, our atoms are located at the center position, and when fully evaporated, are confined to a very small region around it, as shown in Fig. \ref{figMeth1}(b). Thus, for a fixed microwave frequency, we can determine the magnetic field directly from a single image.

During normal operation, we use this tomographic technique to characterize our field $\approx 5$ s after every shot. This allows us to know our field, in spite of slow drifts in our field on the order of 30 mG, with a precision of 3 mG from a single experimental execution. There exists a small but nonzero offset between the calibration and the main experimental sequence, and to control for this offset we precisely measure it each day. No detectable drift between the field calibration and the main experimental sequence exists over the course of a day, as verified from several hundred shots taken over the course of several hours.

The primary systematic uncertainty in the magnetic field is due to the uncertainty in the field curvature. For typical experimental conditions, this contributes approximately 3 mG. In addition, temperatures of 100 nK in both lithium and cesium yield a cloud width in the magnetic trap corresponding to approximately 2 mG, leading to slight asymmetric broadening of any features narrower than this. Finally, as noted in the previous paragraph, we measure the offset between the field calibration and the main experimental sequence daily, averaging the measurement until the uncertainty is less than 1 mG. This calibration uncertainty leads to an additional systematic. Combining all of these systematic uncertainties leads to an estimate of the absolute systematic uncertainty of approximately 4 mG. Even with a more conservative estimate of 10 mG which we use in the Letter, our results are not significantly changed.

While systematic errors due to the field curvature influence the absolute magnetic field positions of measured features, they are common to all measurements performed in our system, and therefore do not influence the relative magnetic field positions of these features. For relative magnetic field measurements, the uncertainty due to the magnetic field curvature and the finite temperature effect are negligible. However, the daily offset measurement uncertainty does remain for the relative magnetic field measurement, given by the combined uncertainty of the two uncorrelated offset measurements, leading to a relative systematic uncertainty of 1.4 mG. Finally, the Feshbach resonance position statistical uncertainty must be included, leading to a total relative uncertainty of 2 mG for the narrow Feshbach resonance and 10 mG for the broad Feshbach resonance.

\noindent\textbf{Experimental sequence.} We load and evaporate Li in the dual color trap and Cs in the elliptical trap in a manner similar to the procedure described in Refs. \cite{Chin4,Chin2}. As the elliptical trap can be translated vertically, Li and Cs are separated for most of the preparation. Preparation ends as Cs is translated to overlap with Li while the beam powers are gradually changed to allow mixing of Li and Cs. At the same time, the field is ramped to a value slightly above the Feshbach resonance to be studied: for the narrow resonance, 850 mG away; for the broad resonance, 5.2 G away. In both cases, the absolute value of the scattering length is below 400 $a_0$, sufficient to suppress interactions during the final stage of preparation.

We then jump the magnetic field to the desired value and hold for an interaction time at fixed magnetic field. The magnetic field is varied from one shot to the next over a fixed range, with the order in which points are taken randomized to eliminate slow drifts. The interaction time is held fixed over each field scan. After the interaction time, the field is jumped back to a fixed value for high field imaging of Li. We then determine the Li and Cs particle number and temperature using absorption imaging.

The sequence differs slightly for different scans. In particular, the cross-thermalization measurement is performed with increased power in the 785 nm beam, as this simultaneously increases the Li trap depth and decreases the Cs trap depth, leading to the required temperature imbalance (T$_\textrm{Li}$=200--400 nK, T$_\textrm{Cs}$=40--50 nK). Furthermore, with the Li at high temperature, we ensure that it does not form a degenerate Fermi gas, avoiding potential many-body effects which would complicate the cross-thermalization measurement. In addition, during this sequence, the Li and Cs clouds are mostly separated. As a result, the collision rate is reduced, allowing for longer interaction times. This is essential to reduce the importance of the magnetic field settling time of $\approx$2 ms. In addition, the three-body recombination rate is suppressed even more than the elastic collision rate by the reduced density in the overlapping region, suppressing three-body loss relative to the cross-thermalization rate.

For the Efimov resonance measurements, a lower 785 nm beam power is used such that the Li and Cs temperatures are approximately equal, with $T\approx 100$ nK. However, for measurements very near the Feshbach resonance, we still use partial overlap to slow the three-body recombination rate, reducing the importance of the magnetic field settling time. For measurements far from the Feshbach resonance, with Li-Cs scattering length $|a|<1000~a_0$, we optimize the overlap between the Li and Cs clouds to compensate for the reduced recombination rate at small scattering length.

After this measurement, we perform a high field calibration, as detailed above. The total time between the end of the main experimental sequence and the field calibration is $\textless$5 s.

\noindent\textbf{Recombination coefficient $K_3$ extraction.} As shown in Fig. 3, raw data consists of the measurement of atom number via absorption imaging after a finite interaction time at a specific magnetic field. In order to compare data across different days and varying conditions, we convert atom loss into a scaled measurement of the three-body recombination coefficient $K_3$, which is normalized to the measured value far away from resonance $K^{(0)}_3$.  

In order to extract the scaled value of $K_3$, we adopt a simple model of loss in our system motivated by a number of observations. First, when Li and Cs are not mixed, we measure a lifetime orders of magnitude longer than when they are mixed. As such, we can neglect both three-body loss due to Cs only collisions and one-body loss. Second, the variation in temperature across each scan is negligible. This allows us to neglect loss due to cross-thermalization and heating atoms out of the trap. With these approximations and assuming an equilibrium distribution, the atom number of Li and Cs can be modeled with the following set of differential equations
\begin{align*}
\frac{dN_{Li}}{dt} &=  -K_3 G(T) N_{Cs}^2 N_{Li} \\ 
\frac{dN_{Cs}}{dt} &=  -2 K_3 G(T) N_{Cs}^2 N_{Li},
\end{align*}
where $G(T)$ is a time-independent geometric factor accounting for the partial overlap of the clouds given by $G(T) = \frac{1}{N_{Cs}^2 N_{Li}} \int n_{Cs}^2(\mathbf{x},T) n_{Li}(\mathbf{x},T) d^3\mathbf{x}$. The factor of two arises from the fact that two Cs atoms are lost in each three-body recombination collision.

As the enhanced (or suppresed) number of atoms lost is small, we find that typically the species with a smaller number of atoms in a given scan exhibits the best signal to noise.  Therefore, when we extract $K_3$ from a given scan, we fit the number of atoms remaining of whichever species shows the largest signal to noise to the above model. 

Due to slight changes of of the trap from day to day, the geometric factor $G(T)$ is difficult to measure to the level of precision necessary.  Therefore, we rescale extracted values of $K_3$ in each scan such that the average values in regions of overlapping magnetic field match across different scans. Further, for clarity in Fig. 4, data are binned according to inverse interspecies scattering length, and presented as the variance weighted mean of all data in each bin.

\newpage

\onecolumngrid

\section*{Supplemental Material}

\beginsupplement

\textbf{Positions of Efimov resonances in various species.} To better contextualize our measurement, we summarize data for a variety of Efimov resonances in both homo- and heteronuclear systems, shown in Tables \ref{tableSupp1} and \ref{tableSupp2}, respectively. This data is also used in Fig. 1 of the Letter. The atomic species, spin states $\ket{F,m_F}$ (where $F$ and $m_F$ denote the total angular momentum and magnetic quantum number, respectively), and magnetic field positions $B_0$ of the Feshbach resonances are included as the necessary parameters to identify the various Feshbach resonances. The van der Waals lengths of the molecular potentials $r_\mathrm{vdW}$ and, for heteronuclear systems, boson-boson scattering lengths $a_\mathrm{B-B}$ are included as values known to determine the Efimov resonance positions from universal theory $a_{th}$. For homonuclear systems, $a_{th}=-9.73r_\mathrm{vdW}$ \cite{Greene1}, while for heteronuclear systems, the prediction is quoted in Table \ref{tableSupp2}. Finally, the resonance strengths $s_\mathrm{res}$ and Efimov resonance positions are included on the Tables. In Table \ref{tableSupp1}, the Efimov resonance position is labeled $a^{(1)}_-$ to indicate that this is the Efimov resonance arising from the first Efimov state. In Table \ref{tableSupp2}, the position is labeled $a_-$ and always refers to the first Efimov resonance, yet due to the suppression of the resonances arising from the first Efimov states near the Feshbach resonances at 889 and 893 G in Li-Cs (see Letter), this is not always the same as $a^{(1)}_-$.

\begin{table}[h]
\caption{First Efimov resonances measured in homonuclear atomic systems.}
\begin{center}
\begin{tabular}{c c c c r r r} 
\hline
Species & $\ket{F,m_F}\,$ & $\,B_0\,$ & $\, r_\mathrm{vdW}$ ($a_0$)$\,$ & $\, a^{(1)}_{-}$ ($a_0$) & $\,s_\mathrm{res}$\\
\hline
$^7$Li & $\ket{1,1}$ & 737.7 & 31.056 & $-252$(10) \cite{Hulet2} & 0.80 \cite{Chin1}\\
$^7$Li & $\ket{1,0}$ & 898.4 & 31.056 & $-264$(11) \cite{Gross1} & 0.58 \cite{Schmidt2012}\\
$^{39}$K & $\ket{1,-1}$ & 33.64 & 64.49 & $-830$(140) \cite{LENS1} & 2.6 \cite{LENS1}\\
$^{39}$K & $\ket{1,-1}$ & 560.72 & 64.49 & $-640$(90) \cite{LENS1} & 2.5 \cite{LENS1}\\
$^{39}$K & $\ket{1,-1}$ & 162.35 & 64.49 & $-730$(120) \cite{LENS1} & 1.1 \cite{LENS1}\\
$^{39}$K & $\ket{1,0}$ & 471.0 & 64.49 & $-640$(100) \cite{LENS1} & 2.8 \cite{LENS1}\\
$^{39}$K & $\ket{1,0}$ & 65.67 & 64.49 & $-950$(250) \cite{LENS1} & 0.14 \cite{LENS1}\\
$^{39}$K & $\ket{1,0}$ & 58.92 & 64.49 & $-950$(150) \cite{LENS1} & 0.11 \cite{LENS1}\\
$^{39}$K & $\ket{1,1}$ & 402.6 & 64.49 & $-690$(40) \cite{LENS1} & 2.8 \cite{LENS1}\\
$^{85}$Rb & $\ket{2,-2}$ & 155.04 & 82.10 & $-759$(6) \cite{Jin1} & 28 \cite{Chin1}\\
$^{133}$Cs & $\ket{3,3}$ & 787 & 101 & $-963$(11) \cite{Grimm2} & 1470 \cite{Grimm3}\\
$^{133}$Cs & $\ket{3,3}$ & -11.7 & 101 & $-872$(22) \cite{Grimm3} & 560 \cite{Grimm3}\\
$^{133}$Cs & $\ket{3,3}$ & 548.8 & 101 & $-1029$(58) \cite{Grimm3} & 170 \cite{Grimm3}\\
$^{133}$Cs & $\ket{3,3}$ & 554.06 & 101 & $-957$(80) \cite{Grimm3} & 0.9 \cite{Grimm3}\vspace{.1mm}\\
\hline
\end{tabular}
\end{center}
\label{tableSupp1}
\end{table}

\begin{table}[h]
\caption{First Efimov resonances measured in heteronuclear atomic systems. Here, $r_\mathrm{vdW,B-X}$ and $r_\mathrm{vdW,B-B}$ indicate the van der Waals lengths of the B-B and B-X intermolecular potentials, respectively, where B represents the identical bosons (in these cases, either Rb or Cs) and X represents the third atom (Li) in the trimer.}
\begin{center}
\begin{tabular}{c c c c c r r r r} 
\hline
Species & $\ket{F,m_F}\,$ & $\,B_0\,$ & $\, r_\mathrm{vdW,B-X}$ ($a_0$)$\,$ & $\, r_\mathrm{vdW,B-B}$ ($a_0$)$\,$ & $\, a_{-}$ ($a_0$) & $\, a_\mathrm{B-B}$ ($a_0$)$\,$ & $\,a_{th}\,$ & $\,s_\mathrm{res}$\\
\hline
$^{7}$Li--$^{87}$Rb & $\ket{1,1}$-$\ket{1,1}$ & 661.44 & 44 & 81.6 & $-1,870$(190) \cite{LiRb} & 100 \cite{LiRb} & $-1,800$ \cite{LiRb} & 3.54 \cite{LiRb}\\
$^{6}$Li--$^{133}$Cs & $\ket{\frac{1}{2},\frac{1}{2}}$-$\ket{3,3}$ & 842.75(3) & 45 & 101 & $-323$(8) \cite{Chin2} & $-1,200$ \cite{Julienne1} & $-330$ \cite{Weidemuller2} & 0.74 \cite{Chin4}\\
$^{6}$Li--$^{133}$Cs & $\ket{\frac{1}{2},\frac{1}{2}}$-$\ket{3,3}$ & 888.577(10) & 45 & 101 & $3,330$(220) & 260 \cite{Julienne1} & $-2,200$ \cite{Yujun1} & 0.02 \cite{Chin4}\\
$^{6}$Li--$^{133}$Cs & $\ket{\frac{1}{2},-\frac{1}{2}}$-$\ket{3,3}$ & 892.648(5) & 45 & 101 & $2,050$(60) & 200 \cite{Julienne1} & $-2,130$ \cite{Weidemuller2} & 0.71 \cite{Chin4}\vspace{.1mm}\\
\hline
\end{tabular}
\end{center}
\label{tableSupp2}
\end{table}

\begin{figure}[h]
\includegraphics[width=3.5 in]{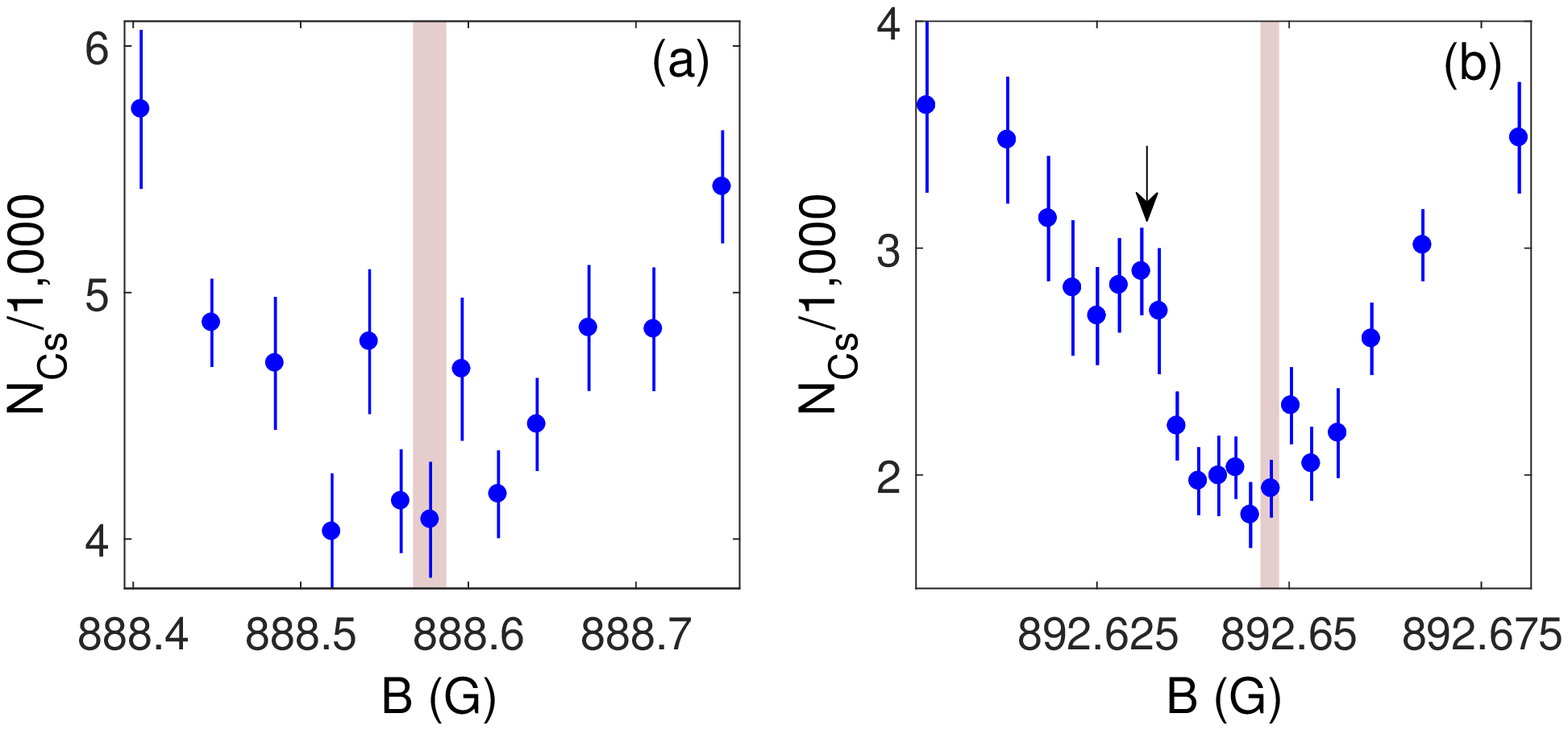}
\caption{Residual three-body loss from the cross-thermalization measurements for the broad (a) and narrow (b) Feshbach resonances. Red shaded regions indicate the Feshbach resonance positions extracted from cross-thermalization measurements. In addition, note the suppression of three-body loss in panel (b) at $\sim 892.63$ G. The arrow indicates the position of this feature based on the fit from the Letter Fig. 3(f).}
\label{figSupp1}
\end{figure}

\textbf{Three-body loss during cross-thermalization.} We show number loss data for Cs in the Feshbach resonance scans in Fig. \ref{figSupp1}. Although not crucial for determination of the Feshbach resonance position, the maxima of three-body loss roughly coincide with the fit position from the cross-thermalization measurement. Furthermore, in Fig. \ref{figSupp1}(b), we note the feature at $\sim 892.63$ G. This position agrees with the fit value for the suppression feature noted in the Letter. The broader context of the Feshbach resonance scan helps to establish that this feature is suppression of three-body loss at $\sim 892.63$ G. This position agrees with that found at positive scattering length, see Letter Fig. 3(f).

\vspace{.01in}

\twocolumngrid

\end{document}